\documentclass[11pt]{article}
\usepackage{eacl2017}
\usepackage{times}
\usepackage{url}
\usepackage{latexsym}
\usepackage{amsmath, amsthm, amssymb, amsfonts}
\usepackage{graphicx}
\usepackage[nottoc]{tocbibind}

\eaclfinalcopy 

\title{Location Inference from Tweets using Grid-based Classification}

\author{Oluwaseun Ajao \\
	SEEECS \\
	Queen's University Belfast \\
	Belfast BT7 1NN, UK \\
	{\tt oajao01@qub.ac.uk} \\\And
	Deepak P \\
	SEEECS \\
	Queen's University Belfast \\
	Belfast BT7 1NN, UK \\
	{\tt deepaksp@acm.org} \\\And
	Jun Hong \\
	SEEECS \\
	Queen's University Belfast \\
	Belfast BT7 1NN, UK \\
	{\tt j.hong@qub.ac.uk} \\}

\date{}

\begin{document}
\maketitle
\begin{abstract}
The impact of social media and its growing association with the sharing of ideas and propagation of messages remains vital in everyday communication.  Twitter is one effective platform for the dissemination of news and stories about recent events happening around the world. It has a continually growing database currently adopted by over 300 million users.  In this paper we propose a novel grid-based approach employing supervised Multinomial Naive Bayes while extracting geographic entities from relevant user descriptions metadata which gives a spatial indication of the user location. To the best of our knowledge our approach is the first to make location inference from tweets using geo-enriched grid-based classification. Our approach performs better than existing baselines achieving more than 57\% accuracy at city-level granularity. In addition we present a novel framework for content-based estimation of user locations by specifying levels of granularity required in pre-defined location grids.
\end{abstract}

\section{Introduction}

The large footprint of Twitter makes it an important marketplace for advertisers to reach its consumers, and projection platforms for government to its citizens. Knowledge of users who interact on Twitter may be quite useful for organisations that render these services. There exists third party domains and other sources such as knowledge bases; These sources amongst others are useful for estimating user locations \newcite{ajao2015survey} however they may be unreliable and insufficient for effectively estimating the location of users. This brings the need to infer locations from transmitted messages solely based on the content alongside other relevant metadata information captured with the tweets such as user description and time zone information etc.

\section{Related Literature}
Location inference also referred to in literature as 'Geolocation Prediction' has enjoyed a fair amount of research interests by several authors working within the space. A few works have been written on the inference of location of Twitter users. The one most related to this work is the \newcite{cheng2010you} as they estimated user locations based solely based on the content of their messages using supervised classification.
\\
Location inference and privacy of geo-spatial data has always been an area of concern \newcite{krumm2007inference} examined the identification of home users from web search data, Privacy still remains a hot-topic of research discussion especially in social media and Twitter in particular \cite{jurgens2015geolocation} \cite{han2016Twitter} with people choosing to hide their online identities to keep an anonymous profile from other users and in some cases for safety and fear of being ‘trolled’ online by cyberbullies
Other works done in the field can be found in Priedosky (2014) who proposed content-based and Gaussian mixture models. \newcite{ikawa2012location} proposed a method that learns association from locations and keywords from previous user messages to predict subsequent messages,  \newcite{jurgens2013s} applied Spatial propagation of location assignments with the knowledge of a few of the locations.
\newcite{compton2014geotagging} inferred location from the friends network with known locations.
\newcite{chang2012phillies} was an unsupervised content-only approach using Gaussian mixture models and maximum likelihood estimation (MLE) 
\newcite{cheng2010you} proposed a probabilistic supervised content-only location inference technique that achieved 51\% accuracy
Also, \newcite{mahmud2014home} used an ensemble of statistical and heuristic classifiers. 
\newcite{ajao2015survey} gave insight into a range of clues for estimating user locations in addition to the message content. 
\newcite{han2014text} used words that gave a spatial clue to indicate the user locations called – location indicative words and information gain ratio-based approach. \newcite{cha2015Twitter} used sparse coding and dictionary learning (PCA whitening, feature augmentation and voting-based grid selection). While in terms of predicting Twitter locations in real-time \newcite{yamaguchi2014online} proposed a solution that constantly infers location of users from the social stream

\section{Methodology}

\subsection{Problem Definition}
The task is in essence inferring user locations on Twitter based on their message content. The problem is defined and stated mathematically thus:\\

Input:\\ 
\begin{enumerate}
   \item Set of users $U = {U_{1}, U_{2}, ... U_{n}}$
   \item Set of tweets T for each user $U_{i}$ as ${t_{i1}, t_{i2}, ... t_{ini}}$
   \item Each tweet $tij = t_{ij}.id, t_{ij}$.UserID, $t_{ij}$.Time, $t_{ij}$.Content, $t_{ij}$.Mentions
   \item Set of labelled Users $U_{L} \subset U$
   \item For each user $U_{i} \in U_{L}, U_{i}.R_{L}$ is the real location of $U_{i}$
   \item Each location $L = [L._{latitude}, L._{longitude}]$
   \item A distance measure $dist(L_{i}, L_{j})$ that gives the distance between $L_{i}$ and $L_{j}$
   \item For Evaluation, for each user $$U_{i} \in (U - U_{L}), U_{i}.R_{L}$$ is the real location of $U_{i}$\\
\end{enumerate} 

Output:\\
For each user $U_{i} \in (U - U_{L}), U_{i}.P_{L}$\\ 
is the predicted location for $U_{i}$
Evaluation Measure to be minimized:\\
$$\sum{Ui\in (U - U_{L})} dist(U_{i}.R_{L}, U_{i}.P_{L})$$

\subsection{Gridding of Geo-tags}
An efficient Java algorithm was employed for the partitioning of the data using the row-major ordering technique as seen in Table \ref{table:grids}.

\begin{figure}[!ht]
  \centering
\includegraphics[width=8.5cm]{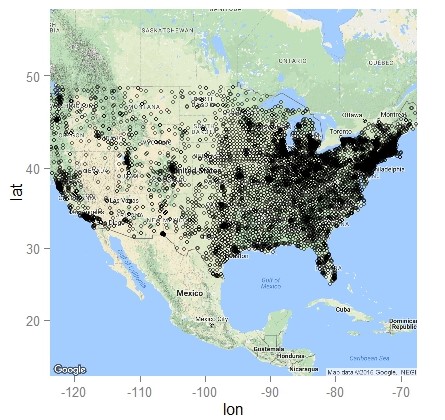}
\caption{US cities with population over 5,000.}
\label{fig:USCities5000}
\end{figure}

\begin{figure}[!ht]
  \centering
\includegraphics[width=8.5cm]{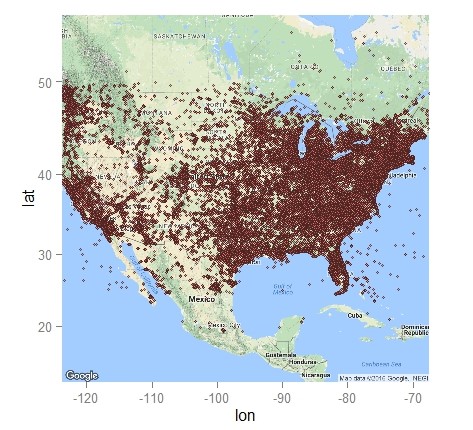}
\caption{Captured tweets geo-located to continental United States.}
\label{fig:CapturedUSCities}
\end{figure}

\begin{table*}
\centering
\begin{tabular}{|c|c|c|c|c|c|c|c|c|c|c|c|c|c|}
\cline{1-4} \cline{6-9} \cline{11-14}
\multicolumn{4}{|c|}{\textbf{8 x 8 Lattice (A)}} &  & \multicolumn{4}{c|}{\textbf{16 x 16 Lattice (B)}} &  & \multicolumn{4}{c|}{\textbf{32 x 32 Lattice (C)}} \\ \cline{1-4} \cline{6-9} \cline{11-14} 
G1        & G2        & ...       & G8       &  & G1         & G2        & ...         & G16       &  & G1        & G2        & ...         & G32        \\ \cline{1-4} \cline{6-9} \cline{11-14} 
G9        & G10       & ...       & G16      &  & G17        & G18       & ...          & G32       &  & G33       & G34       & ...         & G64        \\ \cline{1-4} \cline{6-9} \cline{11-14} 
.         & .         &           & .        &  & .          & .         &          & .         &  & .         &           &          & .          \\ \cline{1-4} \cline{6-9} \cline{11-14} 
.         & .         &           & .        &  & .          & .         &          & .         &  & .         &           &          & .          \\ \cline{1-4} \cline{6-9} \cline{11-14} 
.         & .         &           & .        &  & .          & .         &          & .         &  & .         &           &          & .          \\ \cline{1-4} \cline{6-9} \cline{11-14} 
G57       & G58       & ...       & G64      &  & G241       & G242      & ...      & G256      &  & G993      & G994      & ...      & G1024      \\ \cline{1-4} \cline{6-9} \cline{11-14} 
\end{tabular}
\caption{Grids of Message Geotags}
\label{table:grids}
\end{table*}

\subsection{Supervised Classification}

We  examined a number of machine learning classifiers including the Support Vector Machines (SVM), Logistic regression (Maximum Entropy), Boosting, Bootstrap aggregating (Bagging), Random Forests, Decision trees, Neural Networks and Supervised labelled Dirichlet Allocation (SLDA) and the Multinomial Naive Bayes (MNB) classifier for supervised classification of more than 140,000 geotagged messages that originated from the UK as well as 730,000 messages geotagged to the continental United States. The MNB classifier stood out amongst the lot having the best accuracy and recall. With this preliminary result we then went further to compare our technique with a baseline dataset of 3.7million tweets used by \newcite{chang2012phillies}. In the classification, the words served as the features while the grids served as the labels or predicted results of the task.  In training and testing the classifier, 75\% of the data was randomly split into training while the rest 25\% was used for testing  

\subsection{Data Source}
The UK dataset comprises of the collection of tweets captured in April 2015 from the 4 administrative geolocations in UK. The Twitter API geolocation filters were set to 35 miles radius around Belfast, Cardiff, Glasgow and London. After removal of spam, blank and multiple tweets and blanks using regular expression remained 143,673 tweets. The UK Dataset consisted of 143,673 tweets while the US Dataset of 732,066 geotagged tweets extracted from a larger corpus of 8,663,165 tweets posted from within the US time-zone and filtered with the Twitter streaming API. The following geo-coordinate boundaries were set during the extraction of the messages from Twitter stream 83.162102, -52.23304, 5.49955 and -167.276413

The geo-spatial visualisation of the US tweets is given in Figure \ref{fig:CapturedUSCities}, there's more activity towards the North East of the country; this bears a true resemblance of Figure \ref{fig:USCities5000} which illustrates the population of the United States having 5,000 or more residents according to the 2015 estimates \cite{us2016annual}.

\subsection{Geo-enrichment of messages}
In preparing the US dataset we decided to explore the impact of enriching the messages with geographic entities extracted extracted from their user profiles using algorithm made available by \cite{ritter2011named}an entity recogniser specifically designed for Twitter messages which outperformed the Stanford NER

The following variants were further derived from the tweets:

\begin{itemize}
 \item US Dataset 1: Message text as features and grids as predicted labels
 \item US Dataset 2: Message text + Geographic Entities in User Descriptions as features and grids as predicted labels. 
 \end{itemize}

\subsection{Data Labeling}
The first task was to prepare the training dataset by creating three (3) lattice types were adopted to generate 3 variants of the training datasets. This is based on 8x8, 16x16 and 32x32 gridding approach. We refer to them as Lattices A, B and C respectively. The entire geographic coverage of the tweets were partitioned and based on their coordinates labeled following a row-major ordering as shown in Table \ref{table:grids}. 143,673 pairs of geo-coordinate were automatically assigned grid labels using a Java algorithm. 

\subsection{Data Pre-processing}
The US dataset that was initially captured on the US timeline was of the size The first task was to prepare the training dataset by creating three lattice types were adopted to generate 3 variants of the training datasets. The cleaning of the tweets followed the standard natural language processing pipeline while the Scikit-learn machine learning module in Python for implementing the classifications

\begin{table}
\centering
\begin{tabular}{|c|c|c|c|c|}
\hline
\textbf{Lattice} & \textbf{Precision} & \textbf{Recall} & \textbf{F1} & \textbf{Radius(m)} \\ \hline
8x8              & 0.59               & 0.48            & 0.50        & 120             \\
11x11            & \textbf{0.57}               & 0.49            & 0.42        & 100             \\
16x16            & 0.53               & 0.32            & 0.28        & 60              \\
32x32            & 0.55               & 0.24            & 0.18        & 30              \\ 
\hline
\end{tabular}
\caption{Result of Grid Classification of Baseline Data}
\label{table:baseline}
\end{table}
\section{Analysis \& Results}
Summary results for each grid type, location and classification are presented in Table \ref{table:baseline}. The continental United States has a total geographic area of 6,110,264 square miles \cite{central2016The} gridding of the datasets translated into four (4) variants of radius in miles (120, 100, 60, 30)

\section{Conclusion}
The result of the grid classification shows an improvement over the existing baseline in city-level location inference on Twitter as our method clearly outperforms the existing works. At a radius of 100 miles as show in Table \ref{table:baseline} we achieve an accuracy of 57\% as opposed to the 51\% accuracy achieved by \newcite{cheng2010you} representing a 10\% improvement. Sizes of the lattices can be modified for accuracy and level of granularity or location detail required

\begin{quote}
\begin{verbatim}
\end{verbatim}
\end{quote}

\bibliography{eacl2017}
\bibliographystyle{eacl2017}

\end{document}